\documentclass[aps,prd,amsmath,showpacs,showkeys]{revtex4}
\usepackage{amsmath}
\usepackage[dvips]{graphicx}

\newcommand{\be}{\begin{equation}}
\newcommand{\ee}{\end{equation}}
\newcommand{\bn}{\begin{eqnarray}}
\newcommand{\en}{\end{eqnarray}}
\newcommand{\bd}{\begin{displaymath}}
\newcommand{\ed}{\end{displaymath}}
\newcommand{\bnn}{\begin{eqnarray*}}
\newcommand{\enn}{\end{eqnarray*}}
\def\Ref#1{(\ref{#1})}
\def\Journal#1#2#3#4#5#6{#1, \ #2, \ #3 \  #4 \ (#5) \ #6.}

\begin{document}
\title{Stratification of the phase clouds and statistical effects
of the non-Markovity in chaotic  time series of  human gait for
healthy people and  Parkinson patients}
\author{Renat Yulmetyev} \email{rmy@dtp.ksu.ras.ru}
\author{Sergey Demin}
\author{Natalya Emelyanova}
\author{Fail Gafarov} \affiliation{Department of Physics,
Kazan State Pedagogical University, 420021 Kazan, Mezhlauk Street,
1 Russia}
\author{Peter H\"anggi}\affiliation{Department of
Physics, University of Augsburg, Universit\"atsstrasse 1,D-86135
Augsburg,Germany}
\begin{abstract}
In this work we develop a new  method of diagnosing the nervous
system diseases and a new approach in studying   human gait
dynamics   with the help of the theory  of  discrete non-Markov
random processes \cite{Yulm1}-\cite{Yulm4}. The stratification of
the phase clouds and the statistical non-Markov effects in  the
time series of the dynamics of human gait are considered. We
carried out the comparative  analysis of the data of four age
groups of healthy people: children (from 3 to 10 year olds),
teenagers (from 11 to 14 year oulds), young people (from 21 up to
29 year oulds), elderly persons (from 71 to 77 year olds) and
Parkinson patients. The full data set are analyzed with the help
of the phase portraits of the four dynamic variables, the power
spectra of the  initial time correlation function and the memory
functions of junior orders, the three first points in the spectra
of the statistical non-Markov parameter. The received results
allow to define the predisposition of the probationers to
deflections in the central nervous system caused by Parkinson's
disease. We have found out  distinct differencies between the five
submitted groups. On this basis we offer a new method of
diagnostics and forecasting  Parkinson's disease.
\end{abstract}
\maketitle
\section {Introduction}
A wide class of physical and biological systems exhibits complex
dynamics, related to presence of many factors interacting over a
wide range of time or space scales \cite{Peng1}-\cite{Slat},
\cite{Ehr}-\cite{Jen}. Physiologic signals  generating by complex
self-regulating systems are extemely inhomogeneous, unsteady and
fluctuating in an irregular and chaotic manner. Among them the
locomotion is active movements made by a person which considerably
exceed the norm typical of the size of their body. Walking is the
most widespread form of human locomotion. The study of locomotion
dynamics of a human present great interest for modern biomedicine
and physiology \cite{Haus1}-\cite {Slat}, \cite{Ehr}-\cite{Jen}.
The number of people with various infringements of locomotion
activity increases annually. The basic actual problems connected
with these researches of  the human's gait dynamics and walking
are cosidered in the papers of J. Hausdorff
\cite{Haus6}-\cite{Haus5}. Here, we deal with the changes of the
fractal dynamics of patients' gait \cite{Haus6}, \cite{Haus7}, the
increase of instability of elderly people's gait \cite{Haus6},
\cite{Haus1}, \cite{Haus5} the steady long-range correlation of
fluctuations of young people's step interval  \cite{Haus4} and the
development of children's gait dynamics \cite{Haus8}. As a rule
similar researches are carried out on the basis of the traditional
biomechanical methods \cite{Peng1}, \cite{Peng2}, \cite{Smol},
\cite{Win}, \cite{Holt}, for example on the records of electric
signals in the muscles of legs of a human at walking \cite{Jon}.
Though experimental techniques are well developed at present, the
theoretical analysis of the measurements is inadequate so far.
  The dynamics of the electric signals of
the muscle tremors is related to a class of nonlinear,
nonstationary and nonergodic processes. While analyzing the
electricactivity of the muscle tremors it is necessary to take
into account the discrete non-Markov properties of time change of
the signals as well as the sudden alternation of the behavioral
regimes.  This is the reason why traditional methods of nonlinear
dynamics are not sufficiently sensitive for the purpose of
distinction between different chaotic regimes. The use of the
statistical method of discrete non-Markov random processes
\cite{Yulm1}-\cite{Yulm4} in the similar researches allows to
receive the  more detailed characteristics of deflections  in the
central nervous system (CNS) and in the motor system  and to
differentiate the parameters for  different age groups.

This method also allows to find out the predisposition to
Parkinson's disease as early as possible even of young children.
It is of great practical value as it allows to diagnose and
prevent the development of the disease at its early stages and
reduce the danger of more serious complications, that might set in
without  appropriate treatment.

Parkinson's disease is a chronic one, a slowly progressing disease
caused by the lesion of the extrapyramidal system. The
extrapyramidal system includes the system of nucleus of a brain
and the motor ways which are responsible for  involuntary
automatic regulation and coordination of the complex motor acts,
and also the regulation of a muscular tone, the maintenance of
pose and the organization of impellent manifestation. It is
impossible to establish the cause of the  disease at an early
stage .
 Therefore  the possibility to diagnose  predisposition  to the given disease
 at an  early stage is very important. As a rule the disease
reveals itself  at the age of 50-60, but there are  exceptions.
 160 in 100000 people suffer from this disease.

Parkinsonism is a syndrome of progressing lesion of the nervous
system. This syndrome reveals itself  in the decrease of the
general motor activity, the slowness of movement,  the tremor and
the  increase of  muscular tone. The muscular tone has two
components: a component related to the maintenance of a pose is a
"pose tone",  another  component related to  reflexion is  a
"reflex tone". The disease can be complicated with the  syndrome
of epidemic encephalitis,  atherosclerosis of  brain vessels
caused by  insufficiency of brain blood circulation,   tumors of
 big hemispheres and  craniocerebral traumas. Less often the
disease is connected with  intoxication caused by  manganese,
carbon monoxide, lead and other medical substances products.

The basic external attributes of the disease are reveal themselves
through limitation of the general motor activity. The movements of
the patients are slowed down and performed with difficulty, their
speech is monotonous and hardly audible, the face expression is
unemotional. While walking the patients make short steps, there is
no concomitant motion of their hands and the phenomenon of
propulsia can be observed (inertia movement). The tone of skeletal
muscles increases and as the result the movements are constrained.
The characteristic pose of the patient is as follows: the body is
bent forward, the chin dropped onto the chest, the hands are at
the side of the body bent in elbows. Huntington's disease also
refers to the group under study as it is a certain kind of walking
resembling a dancing gait.

Now more than 10 million people in the advanced countries suffer
from Parkinson's disease. The symptoms of the disease reveal
themselves as tremor and rigidity, caused by mortification of
cells in the certain area of human brain named substantia nigra.
Neurons from this area transfer nerve impulses by means of
doufamin mediator. Owing to the destruction of the nervous cages
of substantia nigra changes take place in other areas of the brain
responsible for motor activity. The condition  of the Parkinson
patient can be alleviated, if the lack of doufamin is filled up.
However the process of neuron destruction continues.

It is supposed, that the "cause" of disease is an unknown toxic
substance contained in the environment. Later on narcotic
substances,  stimulating the phenomena very similar to the
symptoms of Parkinson's disease,  were discovered. Even tea, mint
and herbicides contain substances which  in a combination with the
natural loss of the nervous cages of substantia nigra can cause
Parkinson's disease. The effect of the substantia nigra is thought
to have the effect similar to the effect of toxic substances. It
is connected not only with the destruction of the brain caused by
Parkinson's disease, but cames as a natural "process" of human
ageing.

On April, 16, 2002  "Decode Genetics"  research company has
declared that the define gene is responsible for Parkinson's
disease. If the results of the discovery are put into practice,
the new means of combatting this disease will appear.

It seems very curious and rather odd, that rats and mice are less
sensitive to all the factors listed above, than people and
primates. It is due to the special rodent brain structure.

In Refs. \cite{Haus1}-\cite{Slat} various methods  were offered
for finding the distinction of motor functions of different age
groups and differences between patients and healthy people. In
this work we present a new method of detecting strong differences
existing in different age groups and  the  means of diagnostics
Parkinson's disease at its early stages. A quantitative analysis
of human gait dynamics show, that  the description of motor
processes can not be executed on the basis of the fundamental
methods of statistical physics. The most essential moments in time
series describing the behaviour of real live systems are their
discreteness, nonstationarity, nonergodicity and non-Markovity.

Therefore for the description of the complex systems it is
necessary to use the theory of random non-Markov discrete
processes \cite {Yulm1,Yulm2,Yulm3,Yulm4}. The approach proposed
in these works allows to define a wide spectrum of dynamic
characteristics for the majority of real objects. This method
\cite {Yulm1} is used in the given work for the research of
non-Markov statistical effects in time discrete series of human
gait. The generalization will consist in taking into account the
non-stationarity of stochastic random processes and their further
application to the analysis of human gait by means of time
correlation function (TCF) method.

One of the key moments in the spectral approach in the analysis of
stochastic processes is the use of normalized TCF
\be
a_0 (t) = \frac {\langle {\bf A} (T) {\bf A }(T+t) \rangle}
{\langle | {\bf A (T)} | ^2 \rangle} \, \label {f1}
\ee
where the time $T$ is the beginning of a time serial, the angular
brackets indicate a scalar product of vectors, and vector $\bf A
(t)$ is a state vector of a complex system, $| \bf{A}(t)|$ is the
length of vector $\bf{A}(t)$. The above-stated designation is true
only for stationary systems. In a non-stationary case Eqn.\Ref
{f1} is not true and should be changed. The concept of TCF can be
generalized in case of discrete non-stationary sequence of
signals. For this purpose it is necessary to take advantage of the
standard definition of the correlation coefficient for the two
random signals $X$ and $Y$ in the probability theory
\be
\rho =\frac {\langle {\bf X }{\bf Y} \rangle} {\sigma_X \sigma_Y},
\ \sigma_X = | {\bf X} |, \ \sigma_Y = | {\bf Y} |, \label {f2}
\ee
In Eqn. \Ref {f2} the multicomponent vectors $ {\bf X}, {\bf Y} $
will be generated from fluctuations of signals x and y
accordingly, $ \sigma^2_x, \sigma^2_y $ is proportional to the
dispersion of signals  x  and  y, and values $ | {\bf X} |, | {\bf
Y} | $ represent the lengths of vectors $ {\bf X}, {\bf Y}$,
correspondingly. Therefore the generalization of concept of TCF
\Ref {f1} for non-stationary processes can be served  by the
function \cite{Yulm3}
\be
a (T, t) = \frac {\langle {\bf A} (T) {\bf A} (T+t) \rangle} {|
{\bf A} (T) | | {\bf A} (T+t) |} \ . \label {f3}
\ee
For the quantitative description of non-stationarity with
accordance of Eqns. \Ref {f1}, \Ref {f3} it is convenient to
introduce the function of non-stationarity
 \be
\gamma (T, t) = \frac {| {\bf A} (T+t) |} {| {\bf A} (T) |} =
\left \lbrace \frac {\sigma^2 (T+t)} {\sigma^2 (T)} \right \rbrace
^ {1/2} \, \label {f4}
\ee
which one is equal to the relation of lengths of the vectors of
the final and initial states. In case of stationary process the
dispersion does not vary with the time (or its variations are very
weak).

In this work we consider only a few problems connected with the
description of the human's gait dynamics. The most important of
them are:

1) What are the advantages of using the theory of discreteness and
long-range memory effects for the description of locomotor
dynamics ?

2) Is it possible to use the non-Markov statistical effects for
revealing predisposition to motor system diseases ?

The article is organized as follows. In Section 2 we consider the
brief description of the statistical theory of nonstationary
discrete non-Markov processes in complex systems
\cite{Yulm1,Yulm2,Yulm3}. Section 3 contains the description of
the initial data about electric signals in muscles of legs of
healthy people of various age groups and Parkinson patients. The
received results and their analysis are given in Section 4.
Section 5 is devoted to summary and conclusions of this work.

\section {Statistical theory of nonstationary discrete non-Markov
processes  in complex systems. Basic concepts and definition}

The brief description of the statistical theory of the discrete
non-Markov non-stationary processes for the complex systems of
wildlife is presented in this Section. The theory and the
description of quantities used here can be found in the works of
authors published earlier \cite{Yulm1}-\cite{Yulm4}.

While analyzing complex systems we obtain discrete equidistant
series of experimental data, the so-called random variable
\be
X = \{ x (T), x (T +\tau), x (T+2\tau), \cdots, x (T+k\tau),
\cdots, x (T +\tau N-\tau) \}. \label {f5}
\ee
It corresponds to a measured signal during the time $(N-1) \tau $,
where $ \tau $ is a temporary sampling interval of a signal.

For the dynamical analysis, it is more convenient  to use a
normalized time correlation function (TCF).  For the discrete
processes the TCF has its usual form  ( $t=m\tau, N-1 > m > 1 $ )
\be
a(t)=\frac{1}{(N-m) \sigma(0)\sigma(t)} \sum_{j=0}^{N-1-m}\delta
x(T+j\tau)\delta x(T+(j+m\tau)), \label {f6}
\ee
where $\sigma(0)$ and $\sigma(t)$ is the variances of the initial
$(t=0)$ and final (at moment $t$) states of a systems,
correspondingly.
 The properties of the TCF $a (t) $ are determined by the
conditions of normalization and attenuation of correlations
\be
\lim_{t \to 0} a(t)=1, \lim_{t \to \infty} a(t)=0. \label {f7}
\ee
If we take into account the non-stationarity and discreetness  of
complex systems for real processes, then the kinetic equation for
the TCF $a (t)$ has the form of a closed  set of the
finite-difference equations of the non-Markov type
\cite{Yulm1,Yulm2,Yulm3,Yulm4}:
\be
\frac {\Delta a (t)} {\Delta t} = \lambda_1 a (t) -\tau \Lambda_1
\sum _ {j=0} ^ {m-1} M_1 (j\tau) a (t-j\tau) .\label {f8}
\ee
Here $ \Lambda_1 $ is a relaxation parameter with the dimension of
square of frequency, and parameter $\lambda_1$ describes an
eigen-spectrum of Liouville's quasioperator $ \hat L $
\be
\lambda_1=i\frac {< {\bf A} _k^0 (0) \hat L
 {\bf A}_k^0 (0) >} {< | {\bf A} _k^0 (0) | ^2 >}, ~~ \Lambda_1 =\frac {< {\bf
A}_k^0 \hat L _ {12} \hat L _ {21} {\bf A} _k^0 (0) >} {< | {\bf
A}_k^0 (0) |^2 >}. \label {f9}
\ee
The function $M_1 (j\tau) $ in the rhs of Eq. \Ref {f8} represents
the first memory function
\be
M_1(j \tau)= \frac{< {\bf A}_k^0 (0) \hat L_{12} \{ 1+i\tau \hat L
_ {22} \} ^j \hat L _ {21} {\bf A} _k^0 (0) >} {< {\bf A}_k^0 (0)
\hat L _ {12} \hat L _ {21} {\bf A} _k^0 (0) >}, ~~ M_1 (0)
=1.\label {f10}
\ee
In Eq.\Ref {f9} and later operation $\hat L$ is a
finite-difference operator
\bn
i \hat L = \frac{\bigtriangleup}{\bigtriangleup t},  \ \
\bigtriangleup t = \tau, \nonumber
\en
where $\tau$ is a discretization time  step.

It is easy to see, that in Eq. \Ref {f10} we deal with the time
correlation of new orthogonal dynamic variable $ \hat L _ {21}
{\bf A} _k^0 (0)$.

Eq. \Ref {f8} represents the first equation in the chain of
finite-difference kinetic equations with memory for the discrete
TCF $a (t) $. It is easy to see that the memory function $M_1 (t)
$ takes into account the statistical memory about previous states
of the system. By using the procedure of Gram-Schmidt
orthogonalization \cite{Yulm1} we receive the recurrent formula,
in which the older dynamic variable $ {\bf W} _n = {\bf W} _n (t)
$ is connected to the younger one in the following way:
\be
{ \bf W} _0 = {\bf A}_k^0 (0), ~~{\bf W}_1 = \{ i \hat L-\lambda_1
\}{\bf W}_0, \ldots ~~ {\bf W}_n = \{ i\hat L-\lambda _ {n-1} \}
{\bf W}_{n-1}+ \Lambda_{n-1} {\bf W}_ {n-2}, ~~ n > 1. \label
{f11}
\ee
Introducing the corresponding projection operators we come to the
following  chain of connected non-Markov finite-difference kinetic
equations ($t=m\tau $)
\be
\frac {\Delta M_n (t)} {\Delta t} = \lambda _ {n+1} M_n (t) -\tau
\Lambda _ {n+1} \sum _ {j=0} ^ {m-1} M _ {n+1} (j\tau) M_n (t-j\tau). \\
\label {f12}
\ee
Here $ \lambda _ {n+1} $ is an eigen-value of the Liouville's
quasioperator and the relaxation parameters $ \Lambda _ {n+1} $
are determined as follows
\bn
\lambda_n=i \frac {< {\bf W} _n \hat L {\bf W} _n >} {< | {\bf W}
_n | ^ 2 >}, \ \Lambda_n =-\frac {< {\bf W} _ {n-1} (i \hat L -
\lambda _ {n+1}) {\bf W} _n >} {< | {\bf W} _ {n-1} | ^2 >}.
\nonumber
\en

Eqs. \Ref {f8}-\Ref {f12} are written down in view of that in the
present work we analyze short time series. In this case it is
possible to not take into account the nonstationarity functions
\cite {Yulm3}.

\section {Experimental data and data processing}
In our study  we use the records of electric signals data in legs
muscles at human's walking from the data base of physionet website
\cite {web}. The first group of the data describe the dynamics of
children's gait, aged from 3 to 10, (I type), the second
represents teenagers, 11 to 14, (II type), the third includes
young people, 21 to 29, (III type), the fourth contains the data
on elderly persons, 71 to 77, (IV type), the fifth characterizes
Parkinson patients (V type) \cite {web}. The obtained data were
dealt with the help of the above introduced technique. The set of
the three memory functions was calculated for each sequence of the
data. The power spectra for each of these functions are obtained
by the fast Fourier transform (FFT). Also we will show the phase
portraits in plane projections of the multidimensional space for
the dynamic orthogonal variables. For a more detail diagnostics of
the system we will consider the frequency dependence of the three
first points of the statistical spectrum of the non-Markovity
parameter. In this study we will use the frequency dependence of
the non-Markovity parameter \cite{Yulm1}-\cite{Yulm4}
\be
\varepsilon_i (\omega) = \left\{ \frac {\mu _ {i-1} (\omega)} {
\mu_i (\omega)} \right\}^{\frac {1} {2}}. \label{f13}
\ee
Here $i=1, 2.. $ and $ \mu_i (\omega) $ is a power spectrum of
$i$th level.

For a more detailed analysis of data  we present  a multiplicative
power (MP)on the fixed frequency. This parameter allows to reveal
the precise quantitative distinctions in the power spectra of
different groups of the data
\be
M = \prod_{i=0}^{3} \mu_i(\omega_{spec}),  \  \
 \omega_{spec}=10^{-2}
f.u., \ \  1 f.u.=2\pi/\tau, \label{f14}
\ee
where $\tau$ is  discretization time. In this work the
multiplicative power (MP) is introduced only to account for $
\mu_i (\omega_{spec}) $, where i=0, ... 3. Further it is possible
to use senior memory functions   to reveal a more precise
distinction between the investigated groups of the data.

\section{Discussion of results}
In this section the quantitative and comparative analysis of the
chaotic dynamics of   healthy people's gait and  the gait of
Parkinson patients  will be carried out on the basis of the theory
presented in section III. In Figs. 1-5  the time series of the
initial signals, the  phase portraits of the dynamic variables,
the power spectra of the TCF,  the  junior memory functions and
the frequency dependence of the first three points of non-Markov
parameter are considered.

The time series of the electric signals in leg muscles  for all
the  five groups of the data are given in Fig. 1. The analysis of
the first four orthogonal variables $W_0 $ (Figs.1a), $W_1 $
(Figs.1b), $W_2 $ (Figs.1c), $W_3 $ (Figs.1d) shows  that the
brightest  fluctuations of the dynamic variables can be  observed
in case of Parkinson patients. The time series of the dynamic
variables for the data of III type (young people) have the best
organization. Pronounced fluctuations are observed in the
development of the signals for all the  groups of probationers.
 In Fig. 2 the phase
portraits of the four first dynamic variables in  six plane
projections for the healthy young man (III type) are given. In
this figure the symmetry of the phase clouds about the center of
coordinates is appreciable. In all phase portraits one can see the
centralized nucleus and a few separate points scattered on the
perimeter. The interval of the  dispersal is 0.1 $ \tau $. The
completely different picture is observed in case of the Parkinson
patients (V type) (see, Fig. 3).

In Fig. 3 the symmetry of the phase portraits about the center of
coordinates is visible.  But here another important feature namely
the  stratification of phase clouds can be observed. The
stratification of the phase cloud results in the  uniform
distribution of all the points  all over the space and the
disappearance the well-defined nucleus. The interval of the
dispersion of the points increases up to 0.175 $ \tau $. Such
stratification is typical only of the data of V type. This kind of
 phase clouds corresponds to the condition of  Parkinson's
disease. The analysis of the data of  other groups shows that even
the small stratification of the phase clouds demonstrates the
 failure of the system and makes it possible to predict
Parkinson's disease  at early stages.

It is necessary to note that  certain stratification of the phase
clouds similar to the data of V type is observed for 5 - 6.5 year
old children and for 12 - 13 year old teenagers. This phenomenon
is related to physiological changes of this age. Infringements of
the gait dynamics   are caused by age changes of the brain
regulator functions. Thus it is difficult to predict
predisposition to  parkinsonism at these age periods. In Fig. 4
the power spectra of the TCF $ \mu_0 (\omega) $ and the three
younger memory functions of  $ \mu_i (\omega), i=1, 2, 3 $ for all
five groups of the data are given. The low frequency spectra are
submitted for a more detailed analysis of the data  on a
double-log scale. The spectra for the data of I-IV types have a
specific fractal dependence. The brightest fractality is observed
for the initial TCF $ \mu_0 (\omega) $.

The frequency spectra of Parkinson patients  (V type) considerably
differ from the spectra of the healthy people. In this case the
fractal dependence disappears. Sharp breaks of the linear sites of
spectra on all frequencies are characteristic for all memory
functions. Sharp spectral peaks are especially appreciable on low
frequencies. The similar peaks are also observed  in case of
probationers of the I type, the phenomenon is caused by
insufficiently "mature" coordination and regulation of the
locomotor activity. It is necessary to note, that  the
distribution of power spectra of the memory functions on  high
frequencies for all the groups of the data  is almost identical.
It is easy to see condensation of spectral lines as well as
spectral noise. It means that the most  trustworthy information
can be obtained only on low frequencies.

There are power spectra of the $ \mu_i (\omega _ {spec}), i = 0
... 3 $, where $ \omega _ {spec} =10 ^ {-2} f. u., \ \ 1 f. u. =
2\pi/\tau $, and of the multiplicative power on the fixed
frequency (MP) for the 5 types of probationers in Table 1. The
spectrum of $ \mu_0 (\omega _{spec}) $ presents the greatest
interest. The value of this parameter (in units of $ \tau^2 $) has
the order from $10^1 $ for children and teenagers (I-II type) up
to $10^2 $ for young and elderly persons (III-IV type). For
patients (V type) this parameter is minimal and has the order of
the unit. The increase of this parameter means extension of
long-range memory and  long-range order in the system. The values
of other parameters $ \mu_i (\omega _ {spec}), i = 1, 2, 3 $ for
the data of all the types  are of  order of unit and  are almost
identical.

By analogy with the frequency spectrum of TCF   MP parameter (in
units of $ \tau^2 $) has the order from $10^2 $ for children and
teenagers (I-II type) up to $10^3 $ for young and elderly persons
(III-IV type). For Parkinson patients (V type) this parameter has
the minimal value of the order about $10^1$. The numerical value
MP means the presence of long-range memory and  order in a healthy
system. It is necessary to notice, when the top index in a product
of  $ \mu_i (\omega _ {spec}) (i > 3)$  increases  then the values
of the MP in Eq. \Ref {f14} for different groups get greater
distinctions.

 By analogy it is  possible to use the frequency dependence of the first three points
of the statistical non-Markov spectrum $ \varepsilon_i (\omega) $,
where $i=1,2,3 $ (fig. 5) to  diagnose the diseases of the motor
system. The fractality is most appreciable in the behavior of the
$ \varepsilon_1 (\omega) $ and $ \varepsilon_2 (\omega) $ for I-IV
types of probationers. The spectra of  the third point non-Markov
parameter  $ \varepsilon_3 (\omega) $ for all probationers are
almost identical and take the shape of a straight line $
\varepsilon_3 (\omega)  =1$ with small fluctuations. It means
strong non-Markovity. In frequency spectrum $ \varepsilon_1
(\omega) $ the condensation of the spectral lines
 in the region of 0 < $ \omega $ < 0.2 $ f.u. $ for
I-IV types and their absence for V type is appreciable. For the
probationer of I-III types the spectral discharges close to the
characteristic frequency of 0.2 $ f.u. $ is observed.

The spectral lines for all points of the $ \varepsilon_i (\omega)
$ for V type of probationers take a shape of straight line $
\varepsilon_i (\omega) $ =1 (where  i=1,2,3) with a feebly marked
bursts. For all the groups of the data the behavior of the spectra
of non-Markov parameter $ \varepsilon_i (\omega) $ means the
possibility to describe the dynamics of human gait with the help
of the non-Markovity process with feebly marked splashes of
Markovity on low frequencies.

\begin{flushleft}
Table I

\scriptsize The power spectra $\mu_i(\omega_{spec})$ (in units of
$\tau^{-2}$) and multiplicative power on the fixed frequency (MP)
(in units of $\tau^{-8}$) for all the 5 types (where
$\omega_{spec}=10^{-2} f.u., \ \ 1 f.u.=2\pi/\tau$)
\end{flushleft}
\begin{tabular}{|p{2cm}|p{2cm}|p{2cm}|p{2cm}|p{2cm}|p{2cm}|}
\hline
\ \  & $\mu_0(\omega_{spec})$ & $\mu_1(\omega_{spec})$ & $\mu_2(\omega_{spec})$ &  $\mu_3(\omega_{spec})$  & $M$(MP) \\
\hline\hline

I type & 3 $\cdot 10^1$ & 2 & 1.5 & 2 & 1.8 $\cdot 10^2$ \\
\hline
II type & 4 $\cdot 10^1$ & 2 & 2 & 2 & 3.2 $\cdot 10^2$  \\
\hline
III type  &  $10^2$ & 4 & 2 & 2 & 1.6 $\cdot 10^3$ \\
\hline
IV type  & 2 $\cdot 10^2$ & 7 & 1.5 & 3 & 6.3 $\cdot 10^3$ \\
\hline
V type  & 4 & 4 & 2 & 2 & 6.4 $\cdot 10^1$ \\
\hline\hline
\end{tabular}
\begin{flushleft}
Table II \scriptsize

The three first points of non-Markovity parameter for all the 5
types (when $\omega=0$)
\end{flushleft}
\begin{tabular}{|p{2cm}|p{2cm}|p{2cm}|p{2cm}|}
\hline
\ \  & $\varepsilon_1(0)$ & $\varepsilon_2(0)$ & $\varepsilon_3(0)$  \\
\hline\hline
I type & 4.2 &1.2  &1.1  \\
\hline
II type &3.95  &1.5  & 0.75  \\
\hline
III type  & 4.5 & 1.3 & 0.87  \\
\hline
IV type  & 6.2 & 1.8 & 0.6 \\
\hline
V type  & 1.1 & 1.25 & 0.95  \\
\hline\hline
\end{tabular}
\begin{flushleft}
Table III \scriptsize

The some kinetic and relaxation parameters for the I-V types
\end{flushleft}
\begin{tabular}{|p{2cm}|p{2cm}|p{2cm}|p{2cm}|p{2cm}|p{2cm}|}
\hline
\ \  & $\lambda_1(\tau^{-1})$ & $\lambda_2(\tau^{-1})$ & $\lambda_3(\tau^{-1})$ & $\Lambda_1(\tau^{-2})$ & $\Lambda_2(\tau^{-2})$ \\
\hline\hline
I type (4 y) & -0.29847 & -1.0573 & -1.0403 & -0.02819 & 0.046063  \\
\hline
II type (12 y) & -0.42744  & -1.1251 &-1.0324  &-0.13742  & 0.07488 \\
\hline
III type (26 y)&-0.35072&-1.091&-1.0112&-0.079744&0.047017  \\
\hline
IV type  (75 y)&-0.27593&-1.2192&-1.0646&-0.1436&0.1365  \\
\hline
V type (PD) &-0.91981&-1.0136&-1.0193&-0.19034&0.0062973  \\
\hline\hline
\end{tabular}
\bigskip

In Table 2 the frequency dependence of the first three points of
the non-Markovity parameters $ \varepsilon_i (\omega) $, i=1,2,3
where $ \omega $ = 0 is given. The results for the $ \varepsilon_1
(0) $ present a great interest. The comparative analysis of the
this parameter allows to define the evolution of the dynamics of
human gait. The value of this parameter for the healthy people
varies from 4 up to 7 (according to the age group). The
probationers of type V have the order of a unit. The decrease or
increase of this parameter concerning the average value of $5.5$
for the healthy people means the degree of predisposition to the
diseases of human motor system. The values of others parameters $
\varepsilon_2 (0) $ and $ \varepsilon_3 (0) $ change within an
interval of 1 to 2. This means long-range order and the
statistical memory of the system.

In Table 3 the quantitative data of some kinetic and relaxation
parameters $ \lambda_1 $, $ \lambda_2 $, $ \lambda_3 $, $
\Lambda_1 $ and $ \Lambda_2 $ are given. We notice, that all the
parameters for V type accept the least values.

\section{Conclusions}

In this work the dynamics of human gait is considered as the
random non-Markov process. The statistical theory of discrete
non-Markov processes for real objects and live systems is the best
way to investigate this phenomenon. This theory
\cite{Yulm1,Yulm2,Yulm3,Yulm4} allows to define essential
distinctions between the parameters for healthy people and the
people with infringements of locomotor activity. On the basis of
processing of the experimental data of the electric signals in
legs muscles the time series of the dynamic variables $W_i (t) $
were obtained and memory functions as well as the first three
points of non-Markov parameter were calculated. For the analysis
of the time functions we used the power spectra received by the
FFT. The numerical parameters, given in Tables I-III,  demonstrate
essential distinctions between the five groups of the data (I-V).

It is possible to make the following conclusions based on the
long-range memory conception. There are, at least, the two age
periods (5-6.5 years and 12-13 years), when it is difficult to
diagnose predisposition to the infringement of the locomotor
activity. These periods are connected with the age of
physiological changes in young human organism. Within other age
periods the predisposition to the infringement of the locomotor
activity is possible to be diagnosed. The diseases of such sort
are closely connected to general infringements of human CNS. This
is  very important for preventive diagnostics.

The organization of gait is different for various age groups of
the healthy probationers (I-IV groups) and for the Parkinson
patients (V group). Random movement dynamics is characteristic of
the healthy probationers. Organization and rigidity in gait
dynamics is inherent in the Parkinson patients.

The received results can present practical value in different
studies of other diseases of human motor system (for example,
Huntington's disease) and in diagnosing various infringements of
human CNS.

In this paper we have clearly demonstrated that the set of
relaxation, kinetic and spectral parameters as well as the
characteristics of discrete non-Markov stochastic processes are
valuable for the diagnosis of Parkinson's disease.

Since the similar situation is typical for the majority of the
phenomena in live systems our conclusions is of profound
importance for live sciences.

\section {Acknowledgements}
The authors thank  professor J. Hausdorff for valuable advice and
interesting questions. This work has in part (P.H. and R.Y.) been
supported by the Graduiertenkolleg 283: Nonlinear Problems in
Analysis, Geometry and Physics, of the Deutsche
Forschungsgemeinschaft (DFG), Russian Humanitarian Science Fund
(Grant N 00-06-00005a), Russian Foundation for Basic Research
(Grant N 02-02-16146) and NIOKR RT Foundation (Grant N
06-6.6-98/2001(F)). The authors acknowledge Dr. L.O. Svirina for
technical assistance.

\newpage
\section{Figures Caption}
\begin{figure}[h]
\includegraphics[height=17cm,angle=90]{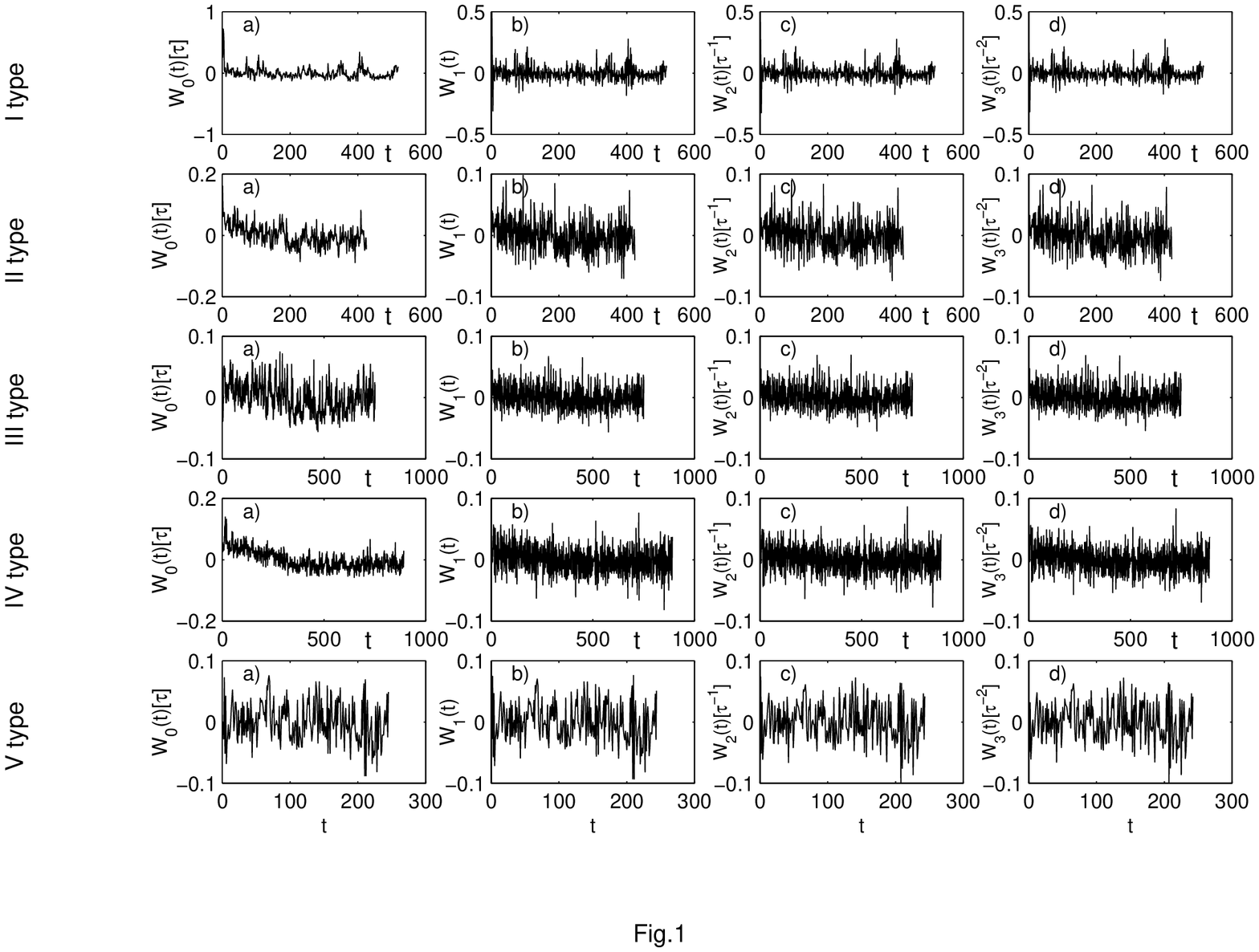}
\caption{The time record of the fourth first orthogonal variables
$W_0 $ (a), $W_1 $ (b), $W_2 $ (c) and $W_3 $ (d) for the child of
4 (I type), for the teenager of 12,4 (II type), for the young man
of 26 (III type), for the elderly person of 75 (IV type), for the
Parkinson patient (V type).}
\end{figure}

\begin{figure}
\includegraphics[height=17cm,angle=90]{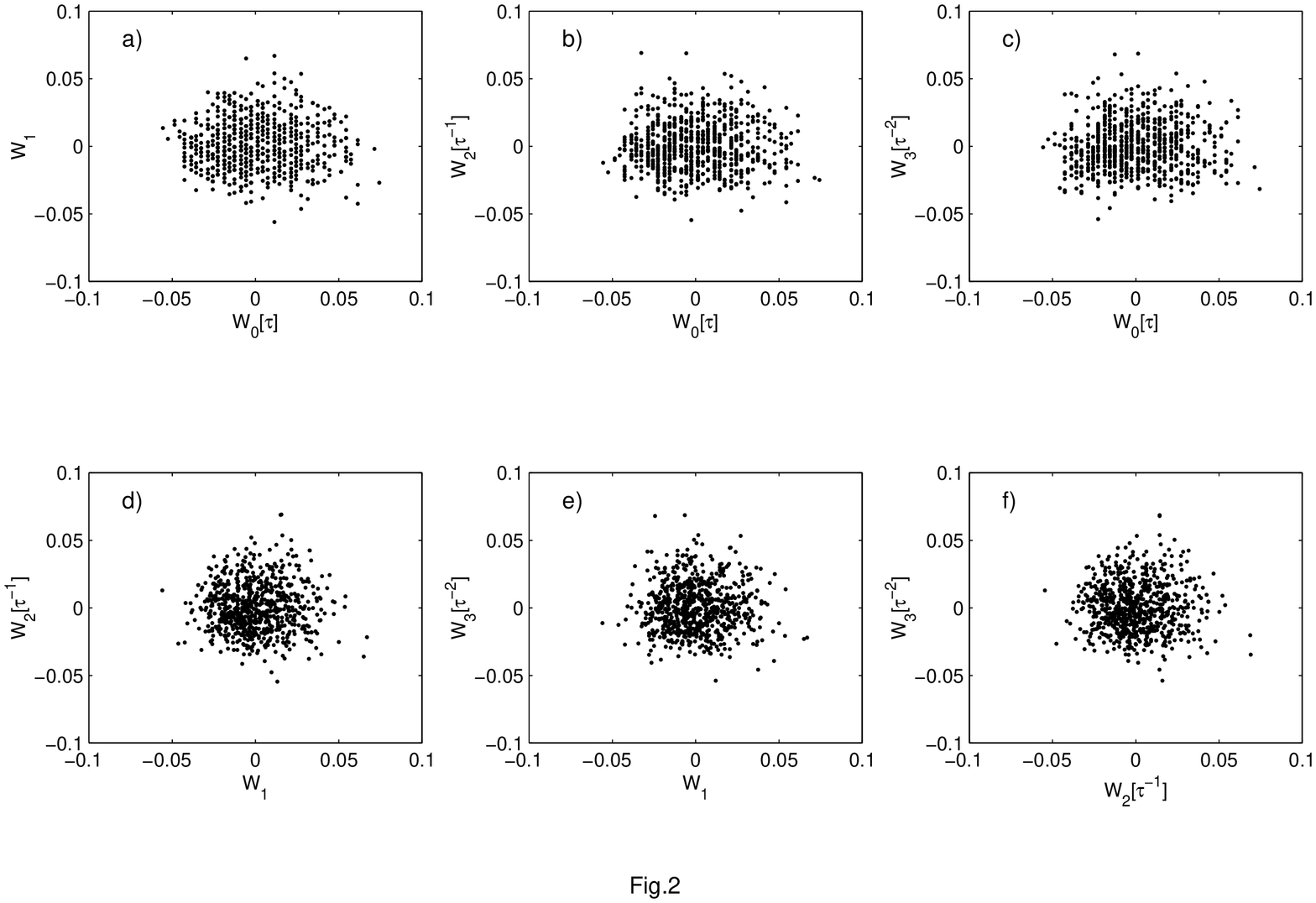}
\caption{The phase portrait of the dynamics of human gait in the
plane projections of two various orthogonal variables ($W_i, W_j$)
for the probationer of type III.}
\end{figure}

\begin{figure}
\includegraphics[height=17cm,angle=90]{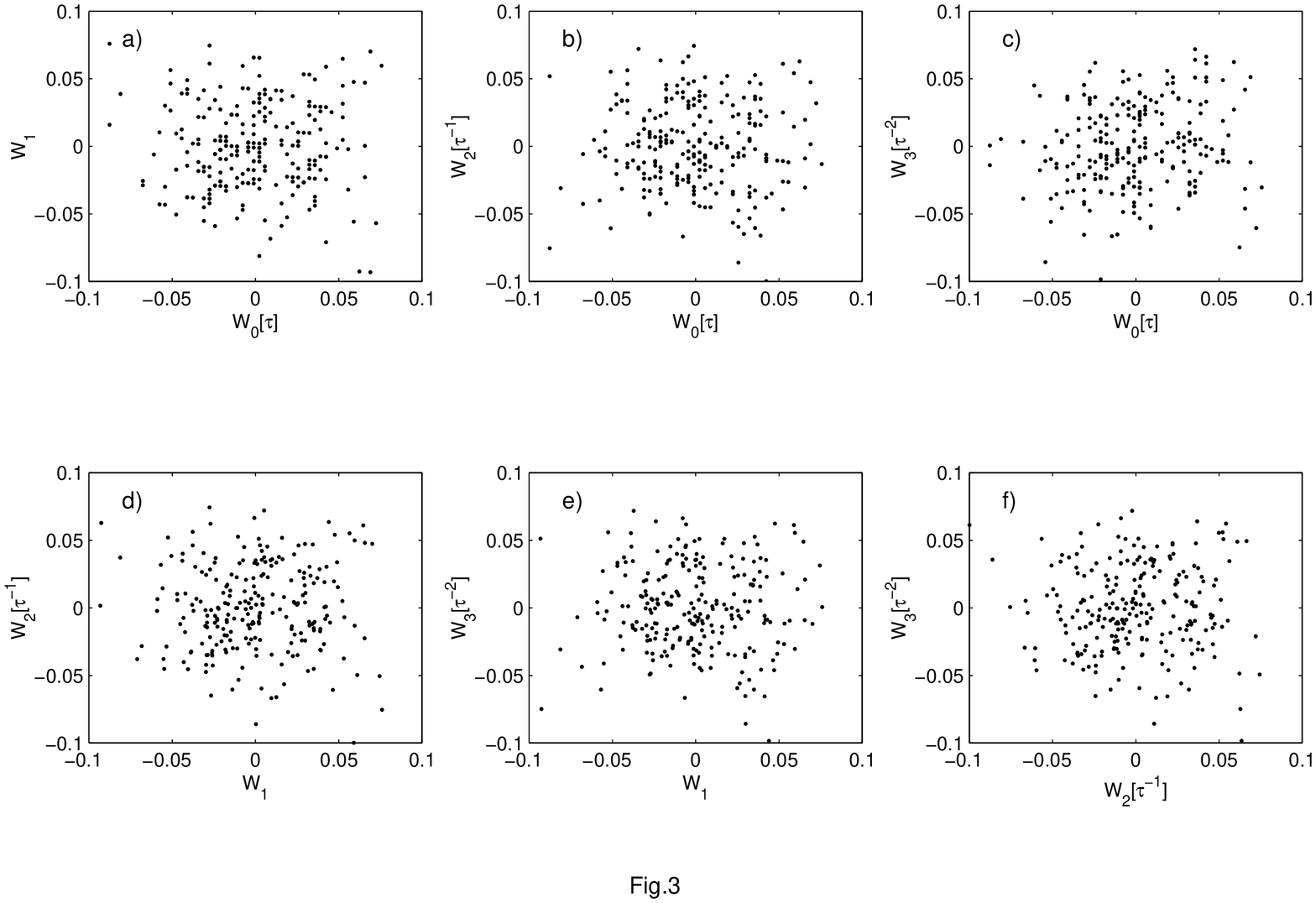}
\caption{The phase portrait of the dynamics of human gait in the
plane projections of two various orthogonal variables ($W_i, W_j$)
for the patient of type V.}
\end{figure}

\begin{figure}
\includegraphics[height=17cm,angle=90]{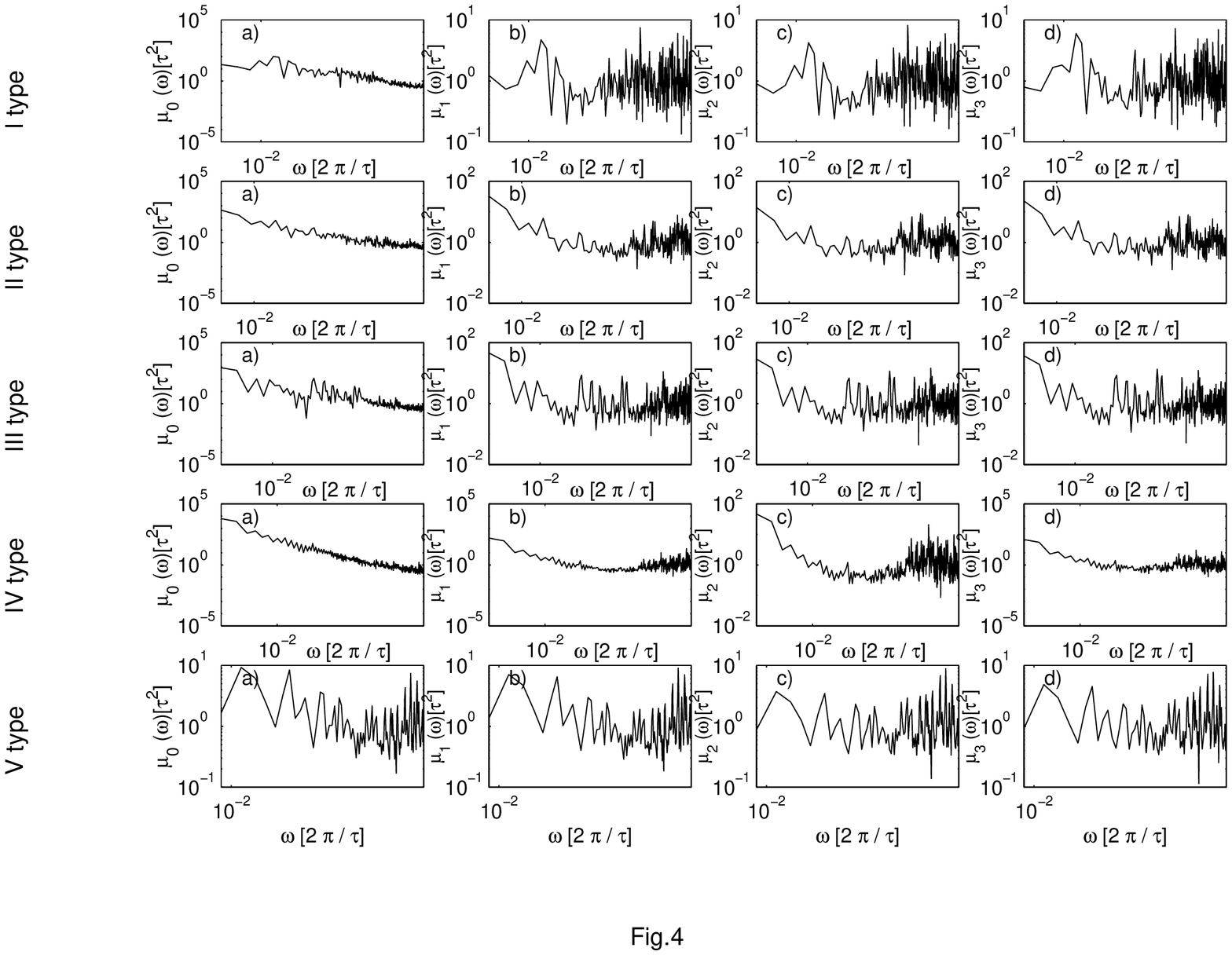}
\caption{The power spectra $\mu_i(\omega)$,  for fourth first
junior memory function of the dynamics of human gait for the all
five types of the probationers ($i=0$ (a), $i=1$ (b), $i=2$ (c),
$i=3$ (d)).}
\end{figure}

\begin{figure}
\includegraphics[height=17cm,angle=90]{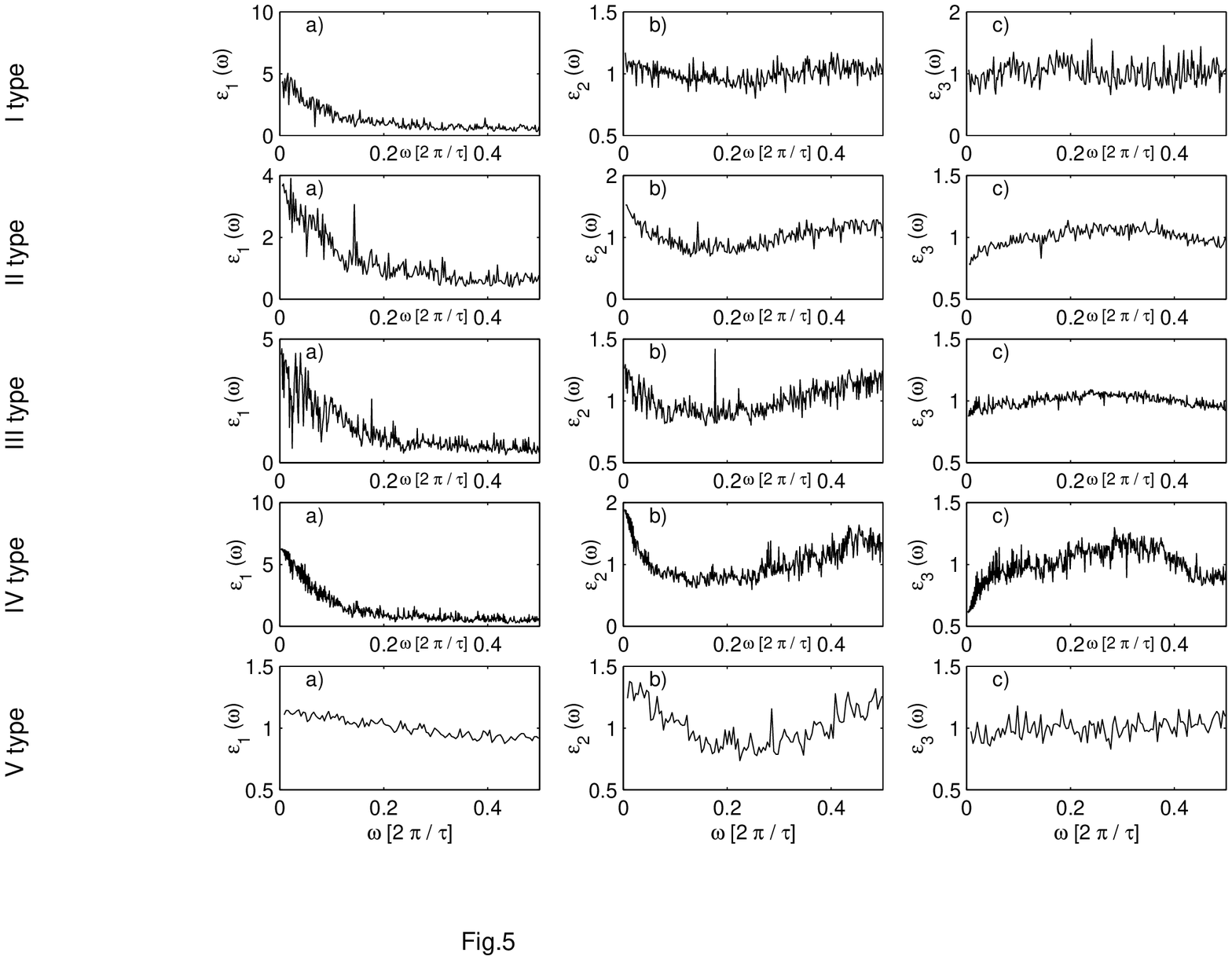}
\caption{The frequency dependence of the first three points of
non-Markovity parameter $\varepsilon_i(\omega)$ of the dynamics of
human gait for all the  5 underlying types (1st point (a), 2nd
point (b), 3rd point (Я)).}
\end{figure}
\end{document}